\documentclass[aps,prb,reprint,showpacs,pdflatex,superscriptaddress]{revtex4-2}
\usepackage{amsmath}
\usepackage{amssymb}
\usepackage{amsthm}
\usepackage{bbm}
\usepackage{graphicx}
\usepackage{hyperref}
\usepackage{physics}
\usepackage{systeme}
\usepackage{times}
\usepackage{epstopdf}
\usepackage{float}

\hypersetup{
  colorlinks=true,linkcolor=blue,citecolor=blue,
  filecolor=blue,urlcolor=blue,breaklinks=true
}

\begin{document}

\title{Effects of curvature on the electronic states of a two-dimensional mesoscopic ring}

\author{Lu\'{i}s Fernando C. Pereira}
\email{luisfernandofisica@hotmail.com}
\affiliation{
        Departamento de F\'{i}sica,
        Universidade Federal do Maranh\~{a}o,\\
        65085-580, S\~{a}o Lu\'{i}s, Maranh\~{a}o, Brazil
      }

\author{Fabiano M. Andrade}
\email{fmandrade@uepg.br}
\affiliation{
        Departamento de Matem\'{a}tica e Estat\'{i}stica,
        Universidade Estadual de Ponta Grossa,\\
        84030-900, Ponta Grossa, Paran\'{a}, Brazil
      }

\author{Cleverson Filgueiras}
\email{cleverson.filgueiras@dfi.ufla.br}
\affiliation{
        Departamento de F\'{i}sica,
        Universidade Federal de Lavras, Caixa Postal 3037,
        37200-000, Lavras, Minas Gerais, Brazil
      }
\affiliation{
  	Departamento de F\'{i}sica,
  	Universidade Federal da Para\'{i}ba, Caixa Postal 5008, 58051-900, Jo\~{a}o Pessoa, Para\'{i}ba, Brazil  }

\author{Edilberto O. Silva}
\email{edilberto.silva@ufma.br}
\affiliation{
        Departamento de F\'{i}sica,
        Universidade Federal do Maranh\~{a}o,\\
        65085-580, S\~{a}o Lu\'{i}s, Maranh\~{a}o, Brazil
      }

\date{\today }

\begin{abstract}
The effects of surface curvature on the motion of electrons in a mesoscopic two-dimensional ring on a cone in the presence of external magnetic fields are examined. The approach follows the thin-layer quantization procedure, which gives rise to a geometry induced potential. Due to the annular geometric shape of the sample, only the mean curvature has relevant effects to the model. Nevertheless, the most significant contribution of the mean curvature occurs in the state $m=0$, which tends to decrease the energies when the magnetic field is null. The effects of curvature are also manifested in the cyclotron frequency as well as in the effective angular momentum through the $\alpha$ parameter, which can be controlled in such a way that the magnitude of these effects becomes explicit. This is verified in the energies and wave functions of the system. A decrease in the number of occupied states in the Fermi energy is observed. As a consequence, there is an alteration in the radial range of the conducting region of the sample. This fact is confirmed by studying the variations in the radii of the states.
\end{abstract}


\maketitle

\section{Introduction}

\label{intro}

Mesoscopic rings have attracted the attention of theoretical and experimental physicists in recent decades for their practicality in revealing various physical phenomena of interest in various branches of physics. It is through these mesoscopic systems that researchers observe the Aharonov-Bohm (AB) effect \cite{PR.1959.115.485}, persistent currents \cite{PLA.1983.7.365,PRL.7.46.1961,PRL.1990.64.2074,PRL.1993.70.2020} and the
quantum Hall effect \cite{PRB.1982.25.2185}. An important contribution to
the physical properties of these devices is the geometric influence on them \cite{PRB.1993.47.9501,PRB.1996.53.6947,SST.1996.11.1635,PRB.1999.60.5626}. In a 1D ring, a thin AB flux tube passes through the center of the ring. The AB flux, however, can be eliminated from the Hamiltonian via a gauge transformation, so that the wave function of an simple electron acquires a
phase. As a result of this operation, the energy and all other derived
physical quantities are periodic functions, with
a period $h/e$. If the flux intensity is increased, a change in the electron
phase arises. In a 2D sample, besides the presence of the AB flux, if we now
add an uniform magnetic field $B$ perpendicular to the flat sample, new
effects are manifested. In this situation, the presence of the magnetic flux
tube will not reveal new effects. As in a $1D$ sample, the energies are
periodic functions of the quantum flux. It is well known that as the magnetic field penetrates the conducting region of the ring, the minimum of the subbands will depend only on the magnitude of the field $B$, and the symmetry that previously existed in the energy profile is then broken. As a
consequence, the energies exhibit an aperiodic behavior.
Moreover, with the presence of the magnetic field, a grouping of electrons into energy sublevels occurs, yielding the Landau levels. As a result, the de Haas-van Alphen (dHvA) oscillations in the magnetization and persistent current are observed. In addition to these oscillations, there are also AB ones. These, however, appear due to the crossing between the energy level curves. The study of dHvA and AB oscillations in rings and quantum dots are issues of fundamental interest in both experimental and theoretical contexts. In Ref. \cite{SST.1996.11.1635}, a exactly soluble theoretical model was proposed to study 2D quantum rings subjected to external magnetic fields. This model is very useful because through it we can get a full quantum mechanical description for the electron states in a 2D ring either threaded by an AB flux tube or subjected to a uniform magnetic field. In recent studies, such a model has been used to investigate other physical systems. For example, to study how the voltage control of the exciton lateral dipole moment induces a transition from singly to doubly connected topology in type-II InAs/GaAsxSb1-x quantum dots \cite{PRAp.2019.11.044011}. In Ref. \cite{PRB.2018.98.205408}, the controlled spin transport through a two-dimensional spin-orbit coupled quantum ring of finite width was studied and the results were used to investigate the magnetoconductance of the ring in which an interface electric field leads to the Rashba mechanism. Another interesting system is the quantum ring configuration of Ref. \cite{OE.2017.25.27857}, where the authors demonstrate the dependence of the emission spectrum of an engineered quantum ring structure on the topological charge of a focused optical vortex driving pulse. In Ref. \cite{PLA.2018.382.432} the authors established an appropriate limit to the confining radial potential of Ref. \cite{SST.1996.11.1635} to investigate the combined influence of the nontrivial topology introduced by a disclination and non inertial effects due to rotation, in the energy levels and the wave functions of a noninteracting electron gas confined to a two-dimensional pseudoharmonic quantum dot, under the influence of an external uniform magnetic field. The influence of both topology and rotation on this model reveals new results, which include a range of magnetic field without corresponding absorption phenomena, which is due to a tripartite term of the Hamiltonian, involving magnetic field, the topological charge of the defect and the rotation frequency. Studies were also performed on quantum information measures in AB ring subject to uniform magnetic fields. It is known that, in order to understand the analysis of quantum information measurements in nanostructures, it is necessary to obtain the exact form of the corresponding one-particle wave \cite{PRB.1995.52.14067,PRB.1999.60.5626,PRB.1996.53.6947}. A treatment involving this issue is presented in Ref. \cite{PLA.2019.383.1110}, where the authors calculate the Shannon quantum information entropies, Fisher informations, Onicescu energies and complexities both in the position and momentum spaces for the azimuthally symmetric two-dimensional nanoring that is placed into the combination of the transverse uniform magnetic field B and the AB flux, and whose potential profile is modelled by the superposition of the quadratic and inverse quadratic dependencies on the radial coordinate.

A few decades ago the engineering of nonplanar mesoscopic objects became a reality. For example, in Refs. \cite{PE.2000.6.828,Nano.2001.12.399}, a technique that made the manufacture of nano-objects of various shapes and sizes possible was developed, while in Ref. \cite{JACS.2005.127.13782}, the synthesis of semiconductor nanocones with contractible apex angles is presented. The possibility of fabricating such objects motivated the research both in the experimental and in the theoretical context to understand the behavior of electrons in these structures when they are in the presence of external fields. Some accomplishments in this direction have been addressed in Refs. \cite{APL.2006.88.212113,PRB.2007.75.205309}. The key question is: how does curvature change the physical properties of a mesoscopic system? When we investigate this question from a theoretical point of view, it is verified that the motion of the particle is rigidly bounded to the surface. As first pointed out by Jensen and Koppe \cite{AoP.1971.63.586} and later perfected by da Costa \cite{PRA.1981.23.1982,PRA.1982.25.2893}, when a quantum particle has its motion constrained to a curved surface, it experiences an effective potential energy whose magnitude depends on the curvature along the surface. Ferrari and Cuoghi \cite{PRL.2008.100.230403} developed a model for studying the motion of a spinless charged particle constrained to move on a curved surface in the presence of an electric and magnetic field. Such a model is a generalization of the one developed by da Costa, which is based on the thin-layer procedure. The main result described in such an approach is that, even in the absence of interactions of any nature, the electrons cannot move around freely on the surface. This implies that we can investigate mesoscopic physical systems in curved geometries by simply controlling the local curvature of the surface and then accessing the physical properties of interest. Subsequently, other generalizations were carried out following the model approached by Ferrari and Cuoghi. Some examples include the derivation of the Pauli equation for a charged spin particle confined to move on a spatially curved surface in an electromagnetic field \cite{PRA.2014.90.042117}, the refinement of the fundamental framework for the thin-layer quantization procedure \cite{AoP.2016.364.68}, the derivation of the formula of the geometric influences of a particle
confined to a curved surface embedded in three-dimensional Euclidean space \cite{PRA.2017.96.022116}, the deduction of the effective equation for a spin-$1/2$ particle confined to a curved surface with the nonrelativistic limit and in the thin-layer quantization formalism \cite{PRA.2018.98.062112} and the inclusion of Dresselhaus hamiltonian with cubic moment to study a nearly free electron system on a curved surface \cite{PRB.2013.87.174413}. In addition to these generalizations, one can find studies on the quantum mechanics of a constrained particle in other contexts. For example, the model of a semiconducting narrow channel with a strong Rashba spin-orbit interaction patterned in a mesoscale serpentine shape \cite{PRB.2018.97.241103}. Basically, this system describes a one-dimensional solid-state electronic setup that operates as a topological charge pump in the complete absence of superimposed oscillating local voltages. Such a device may be a candidate for quantum metrology purposes. In Ref. \cite{PRB.2004.69.195313}, a ring domain in a manifold of negative constant curvature, which is known in the literature as the Lobachevsky plane, was used to study the effect of surface curvature on the magnetic moment and persistent currents. This model also makes it possible to study a quantum dot and some interesting results were observed. In the case of a quantum ring, the surface curvature decreases the spacing between neighboring maxima of de Haas-van Alphen type oscillations of the magnetic moment and decreases the amplitude and period of AB type oscillations while in the quantum dot the surface curvature reduces the level degeneracy at zero magnetic fields. In Ref. \cite{PRL.2014.112.257203} a magnetic energy functional is derived for an arbitrary curved thin shell on the assumption that the magnetostatic effects can be reduced to an effective easy-surface anisotropy. The authors considered the surface of a cone and showed that the effect of the curvature can be treated as the appearance of an effective magnetic field. More recently, we studied the effects of the curvature in the spectrum and in the thermodynamics properties of a quantum dot \cite{doi:10.1002}. It has been observed that the energies of states increase, and that $m = 0$ is not a physical state. This last result modifies the pattern of oscillations in both persistent current and magnetization. In theory of elasticity \cite{Landau1986Elasticity}, the influence of curvature appears in the quantum mechanics of a (quasi-) particle on the surface of a membrane. The potential that controls curvature is an exact solution the elastic membrane shape equation. Such potential can be considered as a quantum potential in the two dimensional Schr\"{o}dinger equation \cite{EJP.2016.38.015405}. This membrane model represents the particular shape of a conformon \cite{JPA.2005.38.6121}. In Ref. \cite{NTch.2016.27.135302}, the one-electron Schr\"{o}dinger equation with open boundary conditions is solved numerically to study the electronic ballistic transport in deformed nanotubes. It is well known that quasi-two-dimensional systems may exhibit curvature. These systems, however, incorporate the three-dimensional influence to their internal properties. In fact, as pointed out by da Costa \cite{PRA.1981.23.1982}, the charged particles moving on a curved surface experience a curvature-dependent potential which greatly influence their dynamics (for some illustrative applications see Ref. \cite{PE.2019.106.200}). In the case of a deformed nanotube, the physical implications due to the curvature of the deformations significantly affect the system dynamics. This fact suggests that such a model can be used in design of nanotube-based electronic devices.

In this paper, we use the Schr\"{o}dinger equation obtained by Ferrari and Cuoghi to study the effects of curvature on an electron in a localized two-dimensional ring in the presence of a uniform magnetic field and a magnetic flux tube. We use the localized ring model proposed by Tan-Inkson. Because it is a very flexible model, both the radius and the width of the ring can be adjusted independently by suitably choosing the two parameters of the confining radial potential, we chose to work with a line element that describes a cone. In addition, since the motion of the electron is constrained to the cone surface, it experiences a geometrical potential, which arises due to the two-dimensional confinement of the particle on the cone surface. The cone, however, has a parameter that controls it. Thus, the model we study in this paper exhibits three parameters that can be adjusted independently. The aim of this paper is to solve the Schr\"{o}dinger equation in this model and determine the energy spectrum and wave functions exactly. We should show how surface curvature affects the electronic structure of the system and describe the most relevant physical implications.

\section{The localized two-dimensional ring}
\label{sec2}

In this section, we briefly review the basic concepts and some results of the localized ring model proposed by Tan-Inkson \cite{SST.1996.11.1635}. This will help us in the understanding of the numerical results that we will provide later. The radial potential that describes the model is given by
\begin{equation}
V\left( r\right) =\frac{a_{1}}{r^{2}}+a_{2}r^{2}-V_{0},  \label{pot-rad}
\end{equation}
where the first term represents a repulsive potential, the second one is a
harmonic oscillator-type potential that constrains the particle to the ring
and $V_{0}=2\sqrt{a_{1}a_{2}}$. This model was proposed to describe a localized
ring of finite width, providing a convenient theoretical tool to study
electronic states as well as their dependence on the magnetic field in a $2D$
ring.

The radial potential (\ref{pot-rad}) has a minimum located at $r=r_{0}=\left(
a_{1}/a_{2}\right) ^{1/4}$, with $r_{0}$ defining the average radius of the
ring. In particular, when $r$ is near $r_{0}$ the potential of the ring has the parabolic form
\begin{equation}
V_{p}\left( r\right) \simeq \frac{1}{2}\mu \omega _{0}^{2}\left( r-r_{0}\right)
^{2},
\end{equation}
where
\begin{equation}
\omega_{0}=\frac{8a_{2}}{\mu} \label{omzero}
\end{equation}
is a quantity that characterizes the strength of the transverse confinement and $\mu$ is the electron effective mass. For a given Fermi energy, $E_{F}$, the width
of the ring is given by
\begin{equation}
\Delta r=r_{+}-r_{-},  \label{width}
\end{equation}
where
\begin{equation}
r_{\pm }=\left( \frac{V_{0}+E_{F}\pm \sqrt{2V_{0}E_{F}+E_{F}^{2}}}{2a_{2}}
\right) ^{\frac{1}{2}}.  \label{ri-re}
\end{equation}
One can estimate the width of the ring by considering a very low Fermi
energy, i.e, $E_{F}\ll V_{0}$, which provides a value
\begin{equation}
\Delta r=\sqrt{\frac{8E_{F}}{\mu \omega _{0}^{2}}}.  \label{width-aprox}
\end{equation}
Moreover, we can control the ``shape" of the potential $V\left( r\right)$
such that both the radius and the width of the ring can be adjusted
independently by suitably choosing $a_{1}$ and $a_{2}$. The radial potential (\ref
{pot-rad}) can also be used to address other physical systems. We list such systems in Table \ref{tab1}.
\begin{table}[h]
	\centering
	\begin{tabular}{cc}
		\hline
		Model &  Requirement \\ \hline
		1D ring & $\omega_{0}\rightarrow \infty $, $r_{0}=$ constant \\
		Straight 2D wire & $\omega _{0}=$ constant, $r_{0}\rightarrow
		\infty $ \\
		Quantum dot & $a_{1}=0$ \\
		Isolated
		anti-dot & $a_{2}=0$ \\
		\hline
	\end{tabular}
	\caption{Summary of the physical models that can be accessed as a particular case of the radial potential (\ref{pot-rad}).}
	\label{tab1}
\end{table}
These particular limits allows us to make a comparison between the electronic states of a $2D$ geometry and other geometries.

\section{Schr\"{o}dinger equation for a particle on a curved surface}
\label{EM}

In this section, we write down the Schr\"{o}dinger equation that govern the motion of a spinless charged particle constrained to move on a curved surface in the presence of an electric and a magnetic field. We follow the work developed by Ferrari and Cuoghi \cite
{PRL.2008.100.230403} and will make some considerations about it.

Using the thin-layer quantization procedure \cite
{PRA.1981.23.1982} and by making a proper choice of the gauge, it was shown
that the surface and the transverse dynamics in the model addressed in Ref. \cite
{PRL.2008.100.230403} are exactly separable. As a result,
the Schr\"{o}dinger equation is decomposed into its normal ($N$) and into its surface
($s$) component, which are given by the following equations,
\begin{equation}
i\hbar \frac{\partial }{\partial t}\chi _{N}=-\frac{\hbar ^{2}}{2M}\partial
_{3}\partial ^{3}\chi _{N}+U_{\lambda }(q^{3})\chi _{N},  \label{motionn}
\end{equation}
and
\begin{align}
i\hbar \frac{\partial }{\partial t}\chi _{s}& =\frac{1}{2M}\Bigg[-\frac{
	\hbar ^{2}}{\sqrt{g}}\partial _{a}\left( \sqrt{g}g^{ab}\partial _{b}\right) +
\frac{ie\hbar }{\sqrt{g}}\partial _{a}\left( \sqrt{g}g^{ab}A_{b}\right)
\notag \\
& +2ie\hbar g^{ab}A_{a}\partial _{b}+e^{2}g^{ab}A_{a}A_{b}+V_{g}+eU\Bigg]
\chi _{s},  \label{motions}
\end{align}
with $a,b=1,2$, where $e$ is the charge of the particle, $A_{j}$ are the
covariant components of the vector potential, $V_{g}$ is the potential due to
the geometry of the surface and $U$ is the electric potential on the
surface. Equation (\ref{motionn}) is just an one-dimensional Schr\"{o}dinger
equation for a spinless particle constrained on $S$ by the normal potential $
V_{\lambda }(q_{3})$. As we are only interested in the dynamics on the
surface, Eq. (\ref{motionn}) will be ignored in our approach. On the other
hand, we can see that the Eq. (\ref{motions}) includes the geometrical
potential $V_{g}$ \cite{PRA.1982.25.2893}, which also has an impact on the electronic properties of the system.

Following  Ref. \cite{AoP.2015.362.739}, we make a connection with the
description of continuous distribution of dislocations and disclinations in
the framework of Riemann-Cartan geometry \cite{AoP.1992.216.1}. In this
description, the particle is bounded to a surface with a disclination
located in the $r=0$\ region. The corresponding metric tensor is defined by
the line element in polar coordinates,
\begin{equation}
ds^{2}=dr^{2}+\alpha ^{2}r^{2}d\theta ^{2},  \label{metric}
\end{equation}
with $r\geq 0$ and $0\leq \theta <2\pi $. For $0<\alpha <1$ (deficit angle),
the metric (\ref{metric}) describes an actual cone, for which $r$ is the distance along the cone from its apex and $\alpha\equiv\sin{\vartheta}$ is given in terms of
its apex angle, $\vartheta$. For $\alpha >1$
(proficit angle), it represents an anti-cone. According to Ref. \cite
{PRA.1981.23.1982}, the geometric potential $V_{g}(r)$, which is a
consequence of a two-dimensional confinement on the surface, is given by
\begin{equation}
V_{g}=-\frac{\hbar ^{2}}{2M}(\mathcal{H}^{2}-\mathcal{K})=-\frac{\hbar ^{2}}{
	8M}(k_{1}-k_{2})^{2},  \label{ptg}
\end{equation}
where $\mathcal{H}$ and $\mathcal{K}$ are the mean and Gaussian curvature of
the surface given respectively by
\begin{align}
\mathcal{H}& =\frac{1}{2}(k_{1}+k_{2})=\frac{1}{2g}
(g_{11}h_{22}+g_{22}h_{11}-2g_{12}h_{12}), \\
\mathcal{K}& =k_{1}k_{2}=\frac{1}{g}\det (h_{ab}),\;\;\;g=\alpha r,
\end{align}
where $k_{1}$ e $k_{2}$ are the principal curvatures and $h_{ab}$ are the
coefficients of the second fundamental form. In the metric (\ref{metric}),
the geometric potential is given by an inverse squared distance potential
and a $\delta $ function potential, which appear naturally in the model and
depend on the type of cone \cite{AoP.2008.323.3150,JMP.2012.53.122106}. For
the cone ($\alpha <1$), it is given by \cite{EPL.2007.80.46002}
\begin{equation}
\mathcal{K}=\left( \frac{1-\alpha }{\alpha }\right) \frac{\delta (r)}{r}
,\;\;\;\mathcal{H}=\frac{\sqrt{1-\alpha ^{2}}}{2\alpha r}.  \label{mean}
\end{equation}
In this case, the geometry induced potential $V_{g}(r)$ reads as
\begin{equation}
V_{g}(r)=-\frac{\hbar ^{2}}{2M}\left[ \frac{(1-\alpha ^{2})}{4\alpha^{2}r^{2}
}-\left( \frac{1-\alpha }{\alpha }\right) \frac{\delta (r)}{r}\right] .
\label{vscone}
\end{equation}

\section{Eigenfunctions and eigenvalues}

The magnetic flux tube in the background space described by the metric (\ref
{metric}) is related to the vector potential as ($\mathbf{\nabla }\cdot
\mathbf{A}=0$, $A_{3}=0$)
\begin{equation}
\mathbf{A=}\frac{1}{2\alpha }Br\mathbf{\varphi }+\frac{l\hbar }{\alpha er}
\mathbf{\varphi },  \label{pot-vet-mag}
\end{equation}
where $l=\Phi /\Phi _{0}$ is the flux parameter and $\Phi _{0}=h/e$. In this
manner, by considering $\psi _{S}=e^{-iEt}\chi _{S}$, the Schr\"{o}dinger
equation (\ref{motions}) becomes
\begin{equation}
H\chi _{s}\left(
\mathbf{r}\right) =E\chi _{s}\left( \mathbf{r}\right),  \label{radialfull}
\end{equation}
where
\begin{align}
H&= -\frac{\hbar ^{2}}{2\mu }\Bigg\{\frac{1}{r}\frac{\partial }{\partial r}
\left( r\frac{\partial }{\partial r}\right) +\frac{1}{\alpha ^{2}r^{2}}
\left( \frac{\partial }{\partial \varphi }+il\right) ^{2}+\frac{1-\alpha ^{2}
}{4\alpha ^{2}r^{2}}  \notag \\
& -\frac{e^{2}B^{2}r^{2}}{4\hbar ^{2}\alpha ^{2}}-i\frac{eB}{\hbar
	c\alpha ^{2}}\left( \frac{\partial }{\partial \varphi }+il\right) -\left(
\frac{1-\alpha }{\alpha }\right) \frac{\delta (r)}{r}\Bigg\}\notag \\
& +\left( \frac{a_{1}}{r^{2}}+a_{2}r^{2}-V_{0}\right).  \label{hamiltonian}
\end{align}
By considering solutions of the form
\begin{equation}
\chi _{s}(r,\varphi )=e^{im\varphi }f_{m}\left( r\right)
,\;\;\;m=...-1,0,1,...,  \label{ansatz}
\end{equation}
Eq. (\ref{radialfull}) results in the following radial equation for a specific quantum number $m$:
\begin{equation}
-\frac{\hbar ^{2}}{2\mu }\frac{1}{r}\frac{d}{dr}\left( r\frac{d}{dr}\right)
f_{m}(r)+V_{efet}f_{m}(r)=E_{m}f_{m}(r),\label{radial_equation}
\end{equation}
where
\begin{equation}
V_{efet}=\frac{\hbar ^{2}}{2\mu }\left( \frac{L^{2}}{r^{2}}+\frac{r^{2}}{
	4\lambda ^{4}}-\kappa ^{2}\right) + \frac{\hbar ^{2}}{2\mu}\left( \frac{1-\alpha }{\alpha }\right) \frac{\delta (r)}{r} \label{V-efetivo}
\end{equation}
is the effective geometry induced potential.  We define the effective angular momentum, the effective cyclotron frequency, the effective magnetic length and a constant parameter, respectively, as
\begin{equation}
L=\sqrt{\left( \frac{m-l}{\alpha }\right) ^{2}+\frac{2\mu a_{1}}{\hbar ^{2}}-
	\frac{1-\alpha ^{2}}{4\alpha ^{2}}}\;,  \label{Mom.Angular}
\end{equation}
\begin{equation}
\omega =\sqrt{\left( \frac{\omega _{c}}{\alpha }\right) ^{2}+\omega _{0}^{2}}\;,  \label{Freq.ciclot.efet}
\end{equation}
\begin{equation}
\lambda =\sqrt{\frac{\hbar }{\omega \mu }}\;,  \label{comp.magnet}
\end{equation}
\begin{equation}
\kappa ^{2}=\frac{\left( m-l\right) \mu \omega _{c}}{\hbar \alpha ^{2}}+
\frac{2\mu V_{0}}{\hbar ^{2}}.  \label{Kappa}
\end{equation}
In Eq. (\ref{Freq.ciclot.efet}), $\omega _{c}=eB/\mu$ is the cyclotron frequency and $\omega_{0}$ is given in Eq. (\ref{omzero}). It is important to emphasize that the presence of the $\delta$ function in Eq. (\ref{radial_equation}), which is a short-range potential, suggests that singular solutions must be taken into account in this approach. According to von Newman's theory of self-adjusted extensions \cite{Book.1975.Reed.II}, the inclusion of the irregular solution of Eq. (\ref{radial_equation}) requires that $\left\vert L\right\vert <1$ \cite{PRD.2012.85.041701,AoP.2013.339.510}. However, for mesoscopic rings, it can be verified that the term $2\mu a_{1}/\hbar^{2}$ in the effective angular moment (\ref{Mom.Angular}) is much larger than $1$, so that there is no value of $L$ belonging to the range required above. Because of this fact, we ignore the $\delta$ function in the radial equation (\ref{radial_equation}) and we consider only the regular solution in the $r=0$ region. This way, the energy and the wavefunction are given, respectively, by
\begin{equation}
E_{n,m}=\left( n+\frac{1}{2}+\frac{L}{2}\right) \hbar \omega -\frac{\left(
	m-l\right) \hbar \omega _{c}}{2\alpha ^{2}}-V_{0},  \label{Enm}
\end{equation}
\begin{align}
\chi_{S}(r,\varphi )&=\frac{1}{\lambda}\sqrt{\frac{\Gamma (L+n+1)}{2^L n!\,\Gamma (L+1)^2}}e^{ -\frac{r^2}{4\lambda^2}}e^{im\varphi }\left(\frac{r}{\lambda}\right)^L \notag \\
&\times \, {_{1}F_{1}}\left(-n, 1+L, \frac{r^2}{2\lambda^2}\right),
\label{funcao.onda}
\end{align}
where $_{1}F_{1}$ is the confluent hypergeometric function \cite{Book.1972.Abramowitz}. The quantum numbers $n$ and $m$ are related to the radial motion and to the angular momentum, respectively. If the model to be studied is a 2D quantum wire, then $n$ is related to the subband index while $m$ refers to the longitudinal motion. The magnetic flux changes from $m$ to $m-l$, which reflects the gauge transformation. The AB flux changes the phase as well as the electron trajectory, as it can be noticed in Eq. (\ref
{funcao.onda}), which results in a non-parabolic dependence on the AB flux
in the energy eigenvalues.  \cite{SST.1996.11.1635}. The effective angular momentum $L$ (Eq. (\ref{Mom.Angular})) is now modified by curvature effects, which is controlled via the $\alpha$ parameter. The third term into the square root in Eq. (\ref{Mom.Angular}) is the modification coming from the mean curvature of the surface. It should also be noted that the effective cyclotron frequency $\omega$ (Eq. (\ref{Freq.ciclot.efet})) and, consequently, the effective magnetic length $\lambda$ (Eq. (\ref{comp.magnet})) are affected by the curvature as well.

\section{The properties of the model}

\subsection{The $2D$ quantum ring in conical space}

In the considered model, the separation between the neighboring subbands is found to be $\hbar\omega
_{0}$ in the absence of magnetic field.  Otherwise, we have $\hbar\omega $, with $\omega$ given in Eq. (\ref{Freq.ciclot.efet}). By using (\ref{Enm}), it is obtained the minimum regardless each subband, that is,
\begin{equation}
m_{0}=\frac{eBr_{0}^{2}}{2\hbar }\sqrt{1-\frac{1}{a_{1}}\frac{\hbar ^{2}}{
		2\mu }\frac{1-\alpha ^{2}}{4\alpha ^{2}}.}  \label{minimo.curvo}
\end{equation}
Obviously, in the absence of magnetic field, it yields $m_{0}=0$.
Therefore, the magnetic field changes the minimum of the subbands. For $\alpha\equiv 1$, the results obtained in the Ref. \cite{SST.1996.11.1635} for a flat sample are recovered.

The effective potential $V_{eff}$ (Eq. (\ref{V-efetivo})) has a minimum value which is defined as being the radius of a state given by
\begin{equation}
r_{n,m}=\left( 2L\right) ^{\frac{1}{2}}\lambda\;.  \label{raio (n,m)  curvo}
\end{equation}
For $m=m_{0}$, we obtain
\begin{equation}
r_{n,m_{0}}=r_{0}\left( 1-\frac{1}{a_{1}}\frac{\hbar ^{2}}{2\mu }\frac{
	1-\alpha ^{2}}{4\alpha ^{2}}\right) ^{\frac{1}{4}}.  \label{raior0}
\end{equation}
In Eq. (\ref{raior0}), there is no dependence on the magnetic field, so that the radius of the state $m_{0}$ is always given by Eq. (\ref{raior0}). If the ring is defined in a flat surface, then the radius of the state referring to the minimum of the sub-bands is
just the average radius of the ring, $r_{0}$. For $\alpha\neq0$, there is a contribution coming from the curvature of the surface. Thus, we can understand the behavior of the electrons in the conducting region of the ring as follows: for $B=0$T, the minimum of all subbands is given by $m_{0}=0$ and, therefore, all the states with $m\neq 0$ have larger radius
in comparison with $r_{n,m_{0}}$. Then, only the radial region $r\geq
r_{n,m_{0}}$ of the ring will be occupied. In the presence of magnetic field, the minimum of all subbands change according to Eq. (\ref{minimo.curvo}). In this way, the states with $\left\vert
m\right\vert <\left\vert m_{0}\right\vert $ have radii $r_{n,m}<r_{n,m_{0}}$, while the states with $\left\vert m\right\vert \geq
\left\vert m_{0}\right\vert $ have the radii given by $r_{n,m}\geq r_{n,m_{0}}$. So, by increasing the magnetic field intensity, more states with $\left\vert m\right\vert <\left\vert m_{0}\right\vert $ are going to occupy the range of radial distribution region of the ring states, $r_{n,m}<r_{n,m_{0}}$. For electrons in the uniform magnetic field region and free of radial confinement potential, the radius of a given state is found to be $r_{n,m}=r_{n,m}=\sqrt{
	2\left\vert m\right\vert}\lambda_{B}$, where $\lambda _{B}=\sqrt{
	\hbar/\left( \omega_{c}\mu \right) }$ is the magnetic length. In a similar manner, we can define the width of a state as the classically allowed region in the effective confinement potential. It is written as
\begin{equation}
d_{n,m}=2\lambda \sqrt{2\left( n+\frac{1}{2}\right) }=2\lambda \sqrt{2n+1}
\label{largura.curvo}
\end{equation}
and has a dependency only on the radial quantum number, $n$.
It can also be seen that both the radius of the state and its width $d_{n,m}$ decrease when the magnetic field or the radial confinement is increased.

\subsection{A 2D quantum wire and an 1D quantum ring}

Here, we analyze the case for $\omega _{0}$ being constant and $r_{0}\rightarrow \infty$ (see Table \ref{tab1}). From the latter, we find that  $a_{1}\rightarrow \infty$. Suppose a $m^{\prime}$ near the minimum of the subbands (\ref{minimo.curvo}), that is,
\begin{equation}
|m^{\prime }|=|m-m_{0}|\ll \frac{2a_{1}\mu }{\hbar ^{2}}\;.  \label{mlinha}
\end{equation}
Thus way, it follows that
\begin{equation}
E_{n,m}=\left( n+\frac{1}{2}\right) \hbar \omega +\frac{\hbar ^{2}}{2\bar{\mu
}}\frac{\left( m^{\prime }\right) ^{2}}{\alpha ^{2}r_{0}^{2}}-\frac{\hbar
	^{2}}{2\mu r_{0}^{2}}\frac{1-\alpha ^{2}}{4\alpha ^{2}},  \label{Enm-2D-F}
\end{equation}
where $\bar{\mu}=\mu \left[ 1+\left( \omega _{c}/\left( \alpha \omega
_{0}\right) \right) ^{2}\right] $, which shows that the effective electron mass is altered by both the magnetic field and the $\alpha$ parameter. By defining $k_{m}=2 \pi m^{\prime } / L$, with $L=2 \pi r_{0}$ being the circumference of the ring \cite{SST.1996.11.1635} and making $\alpha=1$, we recovered the well-known result of a
straight wire with a parabolic confinement \cite{SST.1994.9.1305}.

In order to find the energies of an electron in an 1D quantum ring, we can use the above result, considering $r_{0}$ constant and $\omega _{0}\rightarrow \infty $. We obtain
\begin{equation}
E_{n,m}=\left( n+\frac{1}{2}\right) \hbar \omega _{0}+\frac{\hbar ^{2}}{2\mu
}\frac{\left( m-m_{0}\right) ^{2}}{\alpha ^{2}r_{0}^{2}}-\frac{\hbar ^{2}}{
	2\mu r_{0}^{2}}\frac{1-\alpha ^{2}}{4\alpha ^{2}}  \label{Enm-1D-R}
\end{equation}
For $\alpha\equiv 1$, it is recovered the result found in \cite{SST.1996.11.1635}.

\subsection{The quantum dot}

A quantum dot is obtained by eliminating the potential term $
a_{1}r^{-2}$ from the radial potential model, that is, by making $a_{1}\equiv0$ and $l=0$. Then, using Eq. \ref{Enm} and the condition just mentioned, we arrive at
\begin{equation}
E_{n,m}=\left( n+\frac{1}{2}+\frac{1}{2}\sqrt{\left( \frac{m}{\alpha }
	\right) ^{2}-\frac{1-\alpha ^{2}}{4\alpha ^{2}}}\right) \hbar \omega -\frac{
	m \hbar \omega _{c}}{2\alpha ^{2}}.  \label{Enm-Dot}
\end{equation}
One can see that the geometric potential, given in terms of the mean curvature, does not allows the $m=0$ state to be physical. The effects of surface curvature on the electronic states in a quantum dot was studied in Ref. \cite{doi:10.1002}. If $\alpha\equiv 1$, the energies of quantum dot in a flat surface is recovered \cite{ZP.1928.47.446,RPP.2001.64.701}.

\section{Numerical analysis}

\begin{figure}[th]
	\centering
	\includegraphics[width=0.3\textwidth]{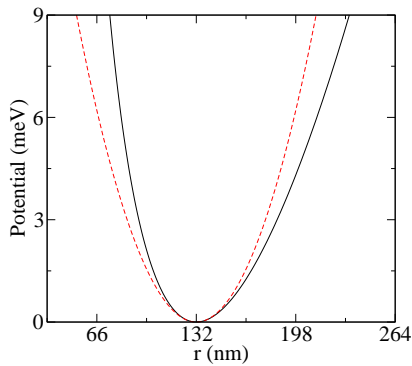}
	\caption{Sketch of the radial potential (solid black line) and the approximate parabolic potential (dashed red line). The potential describes a ring of mean radius $r_{0}=132$ nm and $\hbar \protect\omega_{0}=1.8$ meV.}
	\label{pot-rad-2}
\end{figure}

In the theoretical ring model defined from the binding radial potential
given above, Eq. \ref{pot-rad}, we considered a structure consisted in a GaAs with an
electronic effective mass $\mu=\mu _{e}$, where $\mu _{e}$ is
the rest electron mass. The ring has a average radius $r_{0}=132$ nm e $\hbar \omega_{0}=1.8$ meV \cite{ME.2002.63.47}. In Fig. \ref{pot-rad-2}, we show the sketch of the radial potential using the parameters cited above, making a comparison with the parabolic potential. We can observe that, only for the low energies, the approximate model for the latter is equivalent to the former. At higher energies, the difference between the two models is more pronounced.

\begin{figure}[!h!]
	\centering
	\includegraphics[width=\columnwidth]{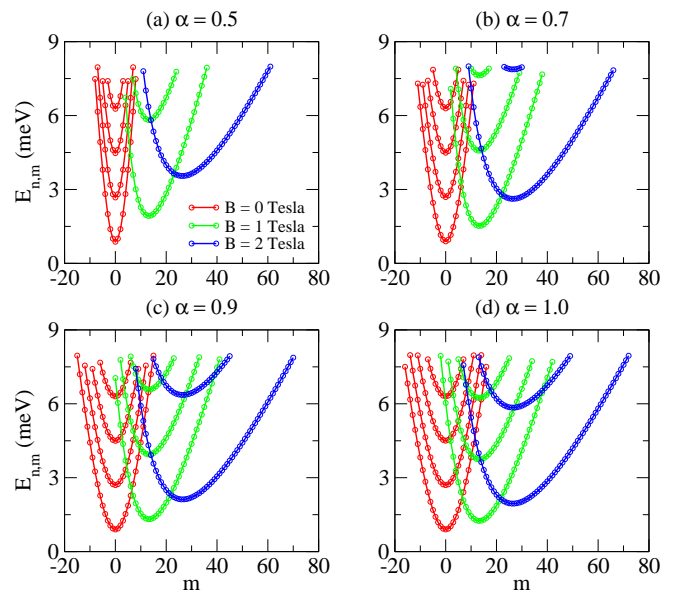}
	\caption{The energies of a $2$D ring as a function of the quantum number $m$ for some values of $\alpha$. We consider a Fermi energy of $E_{F}=8$ meV. The ring is defined with average radius $r_{0}=132$ nm and $\hbar \omega_{0}=1.8$ meV.}
	\label{Enm-m-r0=132nm}
\end{figure}

In Fig. \ref{Enm-m-r0=132nm}, we plot $E_{n,m}$ as a function of the quantum number $m$ for $\alpha =0.5$, $0.7$, $0.9$ and $1.0$, and for some values of magnetic fields. The occupied states are defined in the Fermi energy $E_{F}=8$ meV. The solid lines define the occupied subbands while the circles correspond to the occupied states in each subband. For a null magnetic field, the subbands are symmetric around the minimum of the subbands, $m=0$, because the states are doubly degenerated in its absence. The origin of the double degeneracy is a consequence of the rotational symmetry of the system. Notice that there is no symmetry with respect to the states concerning the different subbands as in the case for the quantum dots \cite{RPP.2001.64.701,Bird.2013}. This symmetry break is caused by the potential term $r^{-2}$ of the radial potential. When the system is subject to curvature effects, the rotational symmetry is preserved. Moreover, we also observed that the neighboring subband separation is given by $\hbar \omega_{0}$
and it does not depend on the curvature. One of the most evident physical
implications in the electronic structure due to curvature is the increase in the energy of states with $m\not=0$ as the $\alpha$ parameter decreases. Namely, this behavior is a consequence of the effect of the curvature on the effective angular momentum of the system for states with $m\not=0$. For $m=0$, the energy of this state tends to decrease because of the effect of the mean curvature, as we can see in Eq. \ref{Enm}.

If a magnetic field is applied, the states configuration change. As we have seen, the minimum of the subbands shifts with the increase of the magnetic field (Eq. (\ref{minimo.curvo})) so that the occupation of the states is altered. As a consequence, a symmetry break with respect to the minimum of the subbands $m_{0}$ occurs.

The separation between the neighboring subbands increases with the presence of the uniform magnetic field. As can be see in Fig. \ref{Enm-m-r0=132nm},
the curvature increases the separation between neighboring subbands, which leads to a decrease in the number of states occupied in the Fermi energy. The effects of the magnetic field in the electronic structure is increased with the presence of curvature.

\begin{figure}[!h!]
	\centering
	\includegraphics[width=\columnwidth]{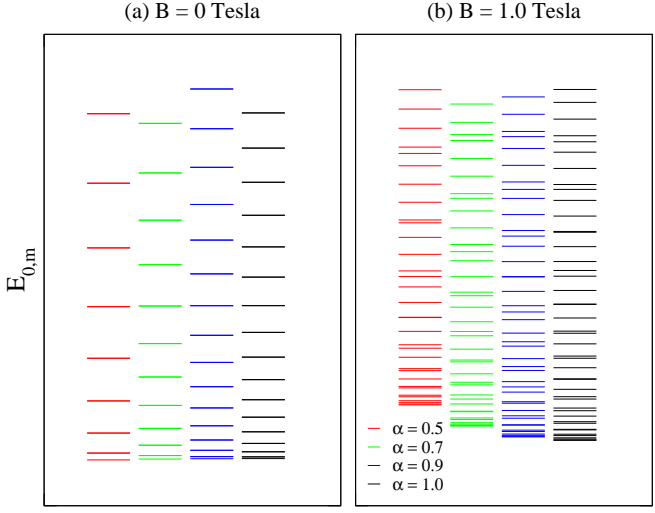}
	\caption{The energy levels for the states of a subband with $n=0$ when $E_{F}=8$ meV (See Fig. \ref{Enm-m-r0=132nm}). For $B=0$ Tesla, each horizontal line corresponds to two energy levels due to symmetry.}
	\label{Enm-niveis}
\end{figure}

In Fig. \ref{Enm-niveis}, we show the subband states with $n=0$ for $B=0$ Tesla and $B=1$ Tesla referring to Fig. \ref{Enm-m-r0=132nm}. Making a comparison between the various values of $\alpha$, we can see that, when the magnetic field is null, the curvature increases the separation between the energy levels so that the number of occupied states decreases. It is also observed that only the state with $m=0$ has a lower energy when $\alpha$ decreases, as already mentioned above.
On the other hand, when the magnetic field is present, a symmetry break occurs. We can see an increase of the minimum energy and a decrease of number of occupied states as $\alpha$ decreases.

\begin{figure}[!h!]
	\centering
	\includegraphics[width=\columnwidth]{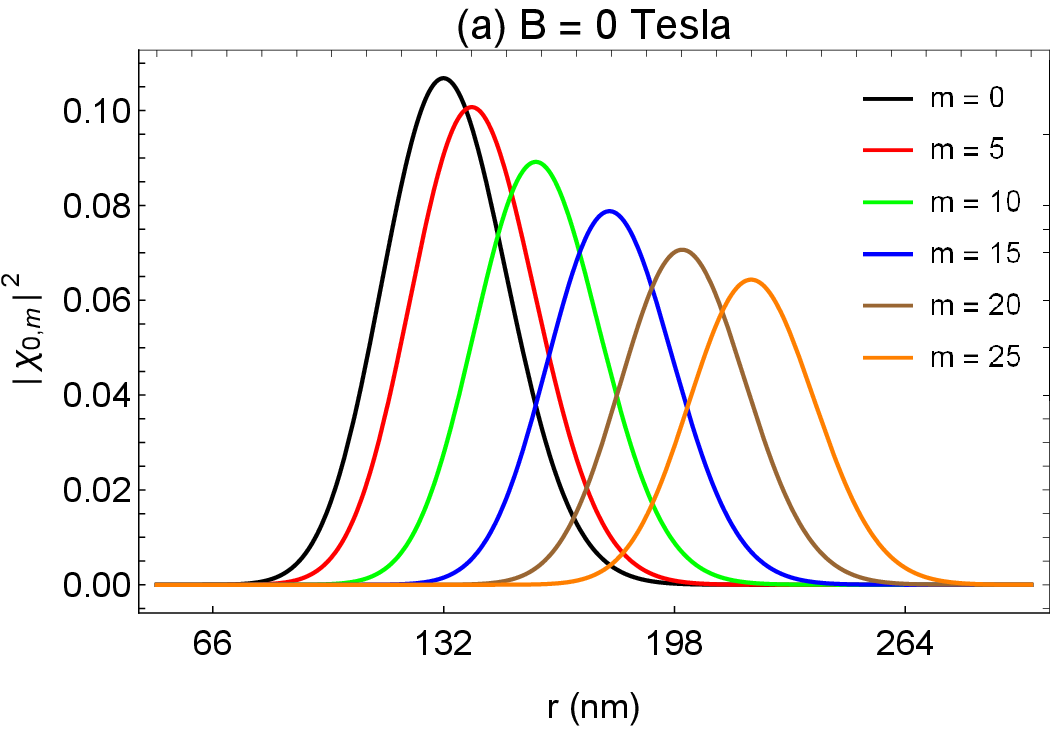}\vspace{0.3cm}
	\includegraphics[width=\columnwidth]{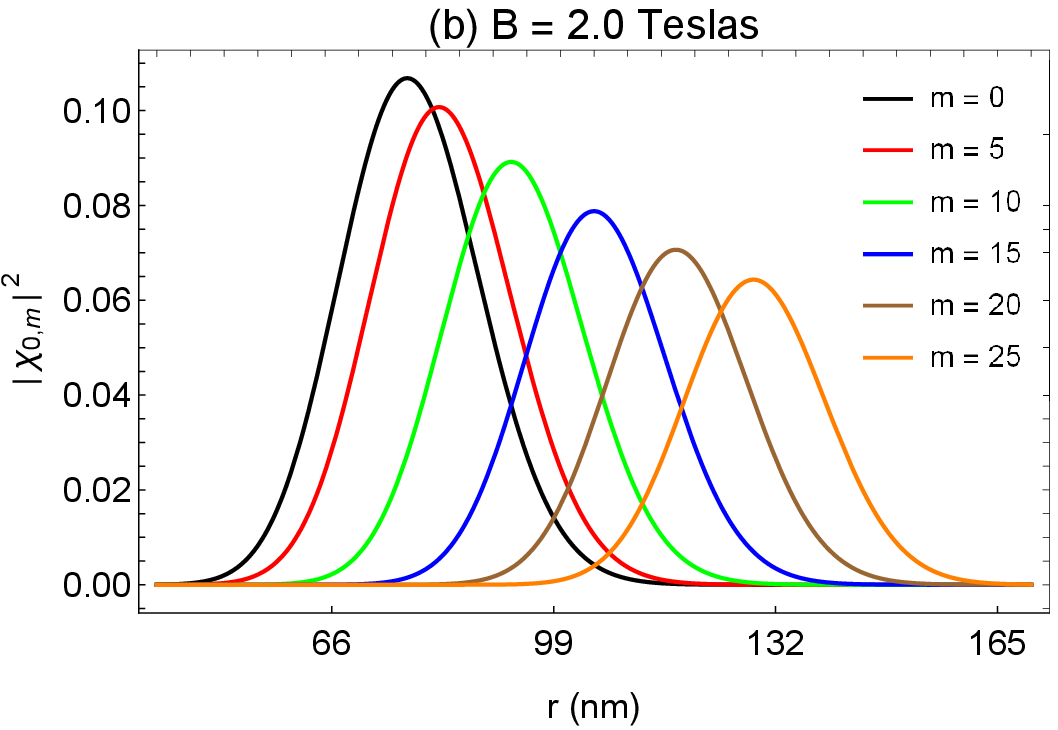}
	\caption{Probability as a function of the radial coordinate for some states of the subband with $n=0$ and $\alpha=0.7$.}
	\label{prob-1}
\end{figure}

\begin{figure}[!h!]
	\centering
	\includegraphics[width=\columnwidth]{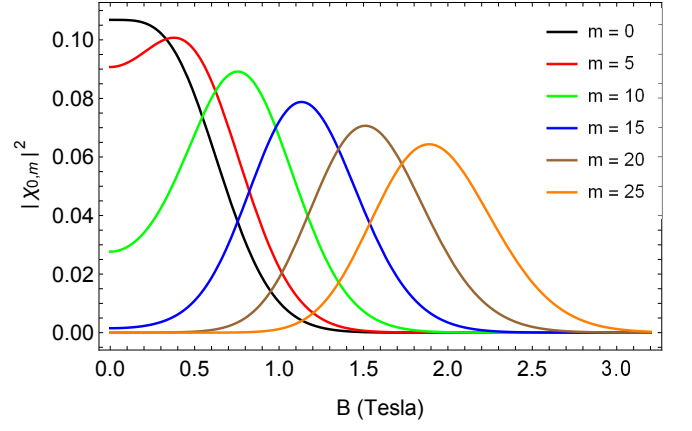}
	\caption{Probability as a function of the magnetic field for the states of the subband with $n=0$ and $\alpha=0.7$.}
	\label{prob-2}
\end{figure}

\begin{figure}[!h!]
	\centering
	\includegraphics[width=\columnwidth]{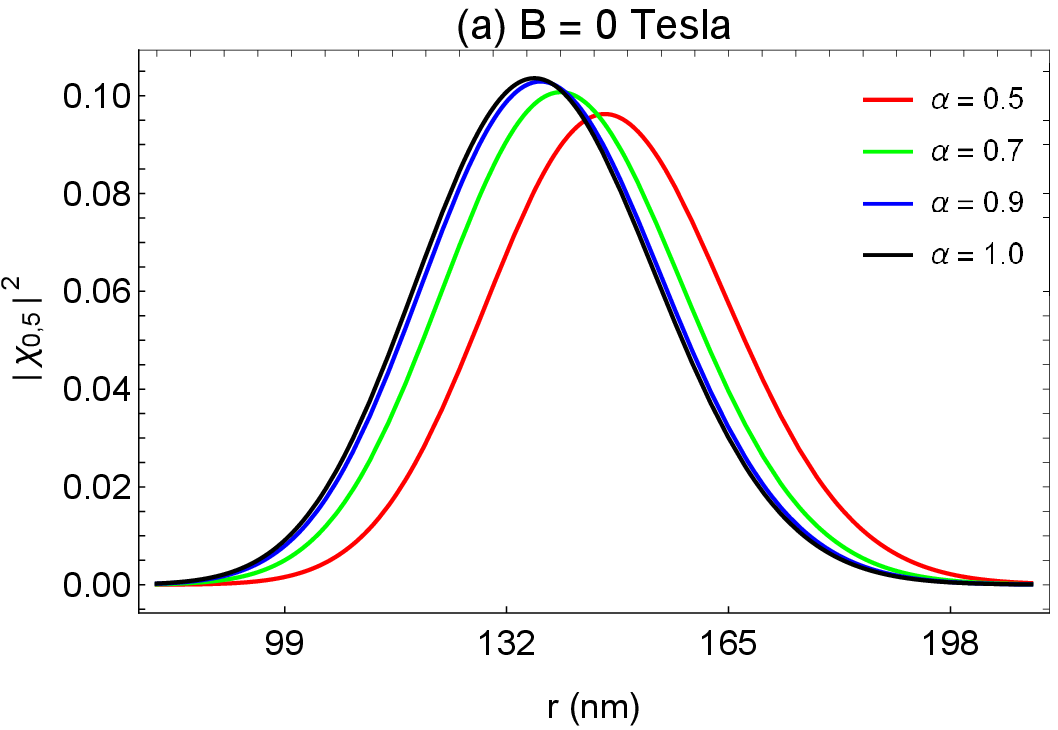}\vspace{0.3cm}
	\includegraphics[width=\columnwidth]{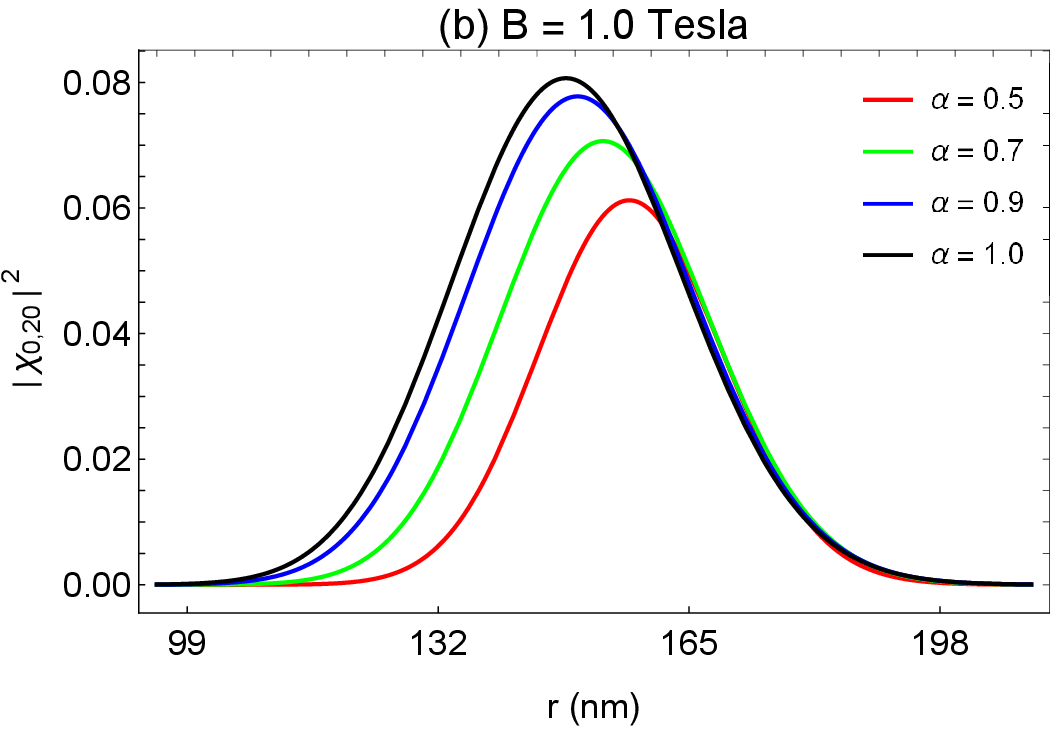}\vspace{0.3cm}
	\includegraphics[width=\columnwidth]{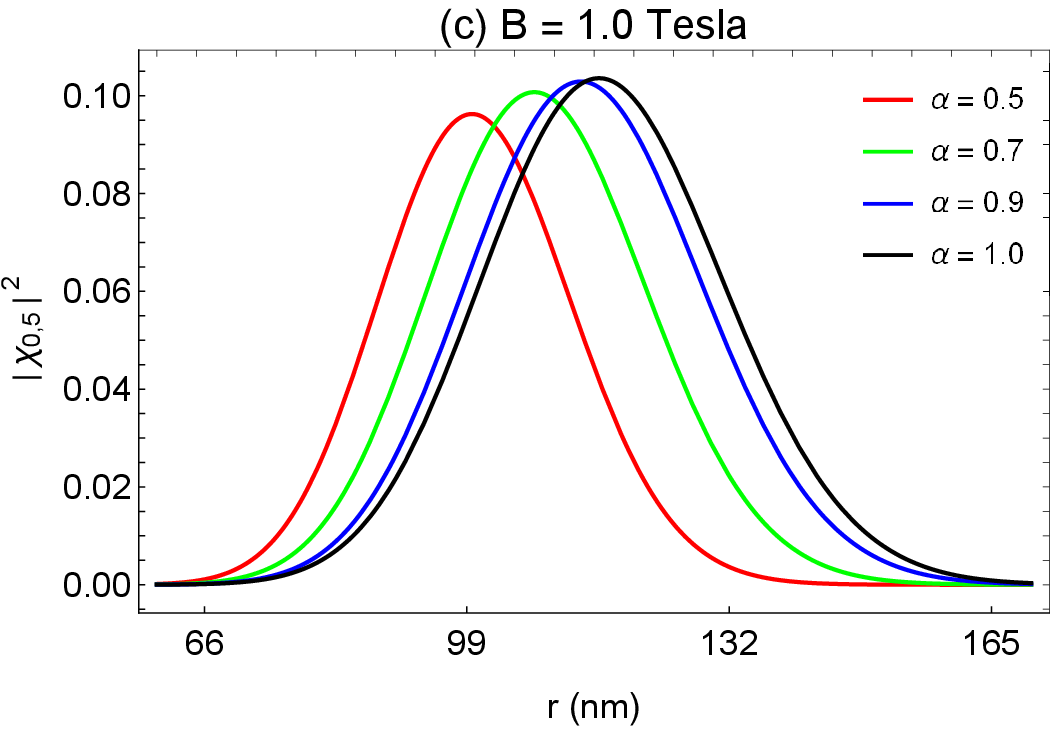}
	\caption{Probability as a function of the radial coordinate. We considered the subband with $n=0$ and some values of the $\alpha$ parameter. In panel (a), we show the profiles of the probability of the state with $m=5$ and a null magnetic field. In panels (b) and (c), we consider the states $m=20$ and $m=5$, respectively, with $B=1$ Tesla.}
	\label{prob-3}
\end{figure}

In the Fig. \ref{prob-1}, we sketch the probability of finding an electron in some states of the first subband for $\alpha=0.7$. When $B=0$ T, the probability amplitude of the state with $m=0$ is maximum for $r_{0}=132$ nm. All states with $\left\vert m\right\vert >m_{0}$ have maximum amplitudes at $r>r_{0}$. Applying a magnetic field, we observe that the maximum amplitude of the states shifts to $r<r_{0}$. Thus, the radii $r_{n,m}$ of the states decrease as the magnetic field increases. This is in accordance with what we have said about the radial distribution of the states in Section $5$.

As shown in Eq. (\ref{minimo.curvo}) and also explicitly in Fig. \ref{Enm-m-r0=132nm}, the minimum of the subbands, changes its state when the magnetic field is varied. So, we can infer that the probability amplitude of a state changes with the magnetic field and, as a result, we observe a maximum probability value when the $m$ state matches the minimum of the sub-bands. In fact, this is observed in Fig. \ref{prob-2}, where we show the probability behavior of some states of the subband with $n=0$ as a function of the magnetic field $B$ for $\alpha=0.7$.

It can also be inferred from the probability that the curvature modifies the radii of the states. In Fig. \ref{prob-3}(a), where we consider the state with $m=5$ and fix the magnetic field at $B=0$T, we see an increase in the radius of the state. In Fig. \ref{prob-3}(b), we fix the value of the magnetic field at $B=1$ T and consider the state with $m=20$, which is to the right of the minimum of the $m_{0}$ subbands. In fact, we can confirm the location of this state by looking at Fig. \ref{Enm-m-r0=132nm}. We see that curvature increases the radius of a state. Finally, in Fig. \ref{prob-3}(c), we keep the magnetic field at $B=1$ T and consider the state with $m=5$, which is to the right of the minimum of the subbands. As can be clearly seen, the curvature decreases the radius of the state. So, if the magnetic field is zero, the radius of the states with $m \neq 0$ increases. However, when the magnetic field is non-null, the states with $\left \vert m \right \vert >\left \vert m_{0} \right \vert $ has its radius increased due to curvature while the radii are reduced for the states with $\left \vert m \right \vert <\left \vert m_{0} \right \vert $. From these analysis, we conclude that by fixing a Fermi energy, the number of states decreases due to curvature, in accordance to the analysis made earlier for the energy as a function of the quantum number $m$.

\begin{figure}[t]
	\centering
	\includegraphics[width=\columnwidth]{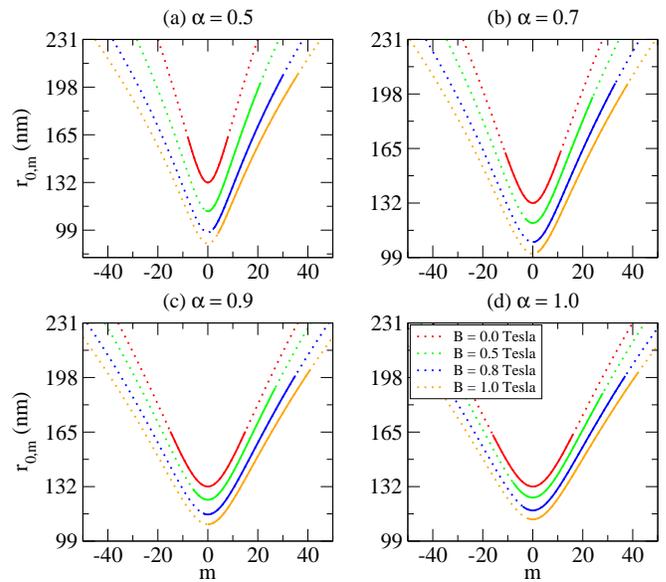}
	\caption{The radius of the state of a $2$D ring (dotted lines) as a function of the quantum number $m$ for some values of $B$ and $\alpha$ parameter. The ring is defined with average radius $r_{0}=132$ nm and $\hbar \omega_{0}=1.8$ meV. The continuous lines represent the occupied conducting region as well as the states that are below Fermi energy, which we take to be $E_{F}=8$ meV.}
	\label{rnm-m-r0=132nm}
\end{figure}

In Fig. \ref{rnm-m-r0=132nm}, we plot the radii of the states $r_{0,m}$ (Eq. (\ref{raio (n,m) curvo})) as a function of $m$ for some values of magnetic field and the
$\alpha$ parameter. We consider a Fermi energy of $8.0$ meV and
look only at the first subband, $n=0$. This allows to observe the occupied
conducting region as well as the interval of occupied $m$ states when the
quantum ring is subject both to the magnetic fields and to the effects of
curvature. The occupied conducting region is the projection of the
continuous line on the vertical axis while the occupied states are projected
on the horizontal axis. As shown above, the radius of a minimum subband is independent of the magnetic field and is always given by Eq. (\ref{minimo.curvo}). For $\alpha=1$, the radius is $r_{0}=132$ nm. Considering $\alpha<1$, this value is slightly changed due to the influence of the mean curvature. When the magnetic field is zero, we saw above that all states with $\left\vert m\right\vert >\left\vert m_{0}\right\vert $ have radii greater than $r_{0,m}=132$ nm, such that only the external conducting region of the ring is occupied. This same situation is observed in Fig. \ref{rnm-m-r0=132nm}. The curvature only decreases the number of states. However, its width does not have significant variations. When $B\neq0$, the radii of the states decreases, so that the internal conducting region of the ring is occupied. The role of curvature is to make the width of the conducting region of the ring quickly populate. However, as the subband is quickly depopulated with the curvature, the width of the conducting region will also decrease more rapidly.
In the Table \ref{tabela-1}, we show the values of $m_{min}$, $m_{max}$, $r_{min}$ and $r_{max}$ referring to the data of Fig. \ref{rnm-m-r0=132nm}. Note that the values of
$r_{min}$, obtained when the magnetic field is zero, are exactly the radii
of the states $m_{0}$ given by Eq. (\ref{raior0}).

\begin{table}[htbp]
	\centering
	\setlength{\arrayrulewidth}{2\arrayrulewidth}
	\begin{tabular}{|c|c|c|c|c|c|}
		\hline
		\textbf{$B (T)$} & \textbf{$\alpha$} & \textbf{$m_{min}$} & \textbf{$m_{max}$
		} & \textbf{$r_{min}$} & \textbf{$r_{max}$} \\ \hline
		0 & 0.5 & -8 & 8 & 131.87 & 163.31 \\
		& 0.7 & -11 & 11 & 131.95 & 162.51 \\
		& 0.9 & -15 & 15 & 132.00 & 165.33 \\
		& 1.0 & -16 & 16 & 132.00 & 163.37 \\ \hline
		0.5 & 0.5 & -1 & 21 & 112.00 & 200.73 \\
		& 0.7 & -3 & 24 & 119.84 & 196.24 \\
		& 0.9 & -6 & 27 & 123.98 & 191.87 \\
		& 1.0 & -8 & 28 & 125.33 & 188.57 \\ \hline
		0.8 & 0.5 & 2 & 30 & 99.40 & 206.02 \\
		& 0.7 & 0 & 33 & 108.30 & 204.48 \\
		& 0.9 & -2 & 35 & 115.12 & 199.16 \\
		& 1.0 & -4 & 37 & 117.55 & 198.94 \\ \hline
		1.0 & 0.5 & 4 & 36 & 96.40 & 206.86 \\
		& 0.7 & 2 & 38 & 102.36 & 204.21 \\
		& 0.9 & 0 & 41 & 109.157 & 202.83 \\
		& 1.0 & -2 & 42 & 112.11 & 200.75 \\ \hline
	\end{tabular}
	\caption{Values of the width of the occupied conducting region of the ring and the interval of states occupied in the
		smaller subband. The data refer to those of Fig. \ref{rnm-m-r0=132nm} for a ring with an average radius $r_{0}=132$ nm and $\hbar\protect\omega_{0}=1.8$ meV.}
	\label{tabela-1}
\end{table}

\begin{figure}[!t]
	\centering
	\includegraphics[width=\columnwidth]{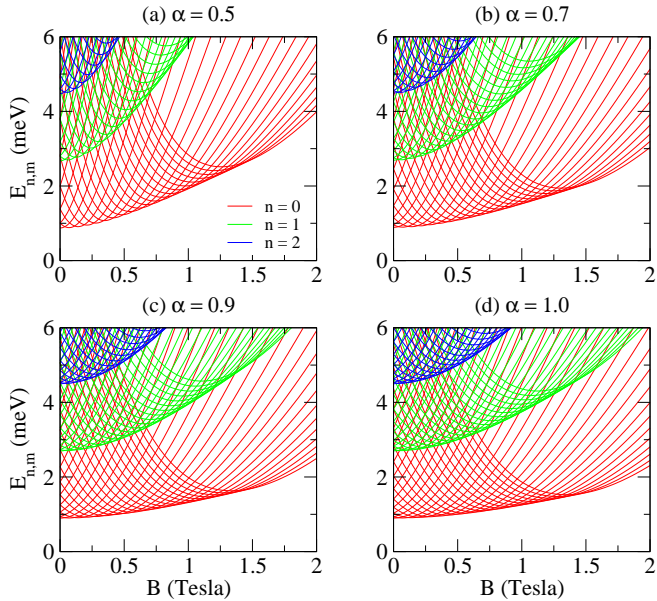}
	\caption{The energy eigenvalues of a $2$D ring as a function of magnetic field B for some values of $\alpha$. The ring is defined with an average radius $r_{0}=132$ nm and $\hbar \protect\omega _{0}=1.8$ meV.}
	\label{E-B-r0=132nm}
\end{figure}

\begin{figure}[!t]
	\centering
	\includegraphics[width=\columnwidth]{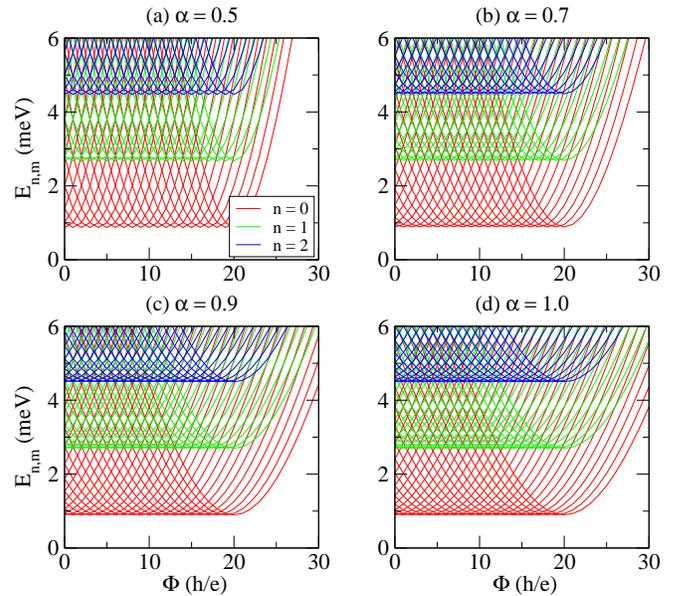}
	\caption{The energy eigenvalues of a $2$D ring as a function of the AB magnetic flux for some values of $\alpha$. The ring is defined with an average radius $r_{0}=132$ nm and $\hbar \protect\omega _{0}=1.8$ meV.}
	\label{E-AB-r0=132nm}
\end{figure}

In Fig. \ref{E-B-r0=132nm}, we show the behavior of the energy states as a function of the magnetic field. We can observe an aperiodic behavior of them, which is more evident when there are several occupied subbands. This behavior is a consequence of the penetration of the magnetic field in the conducting region of the ring \cite{SST.1996.11.1635}. In Figs. \ref{E-B-r0=132nm}(a)-(c), it can be observed an increase in the energy levels due to curvature, which in turn increases the effects of the magnetic field and also modify the effective angular momentum, causing a state to be quickly more unoccupied in the Fermi energy. When the energy spectrum is given as a function of the Aharonov-Bohm flux, a periodic behavior is observed, in contrast to the previous case. As we can see in Fig. \ref{E-AB-r0=132nm}, the energy states are given by a set of translated quasi-parabolas, each one with its center located in the $m=l$ region.

\section{Conclusions}

In this article, we have analyzed the electronic properties of a two-dimensional mesoscopic sample with annular geometry under the effects of curvature and external magnetic fields. The thin-layer quantization procedure has been employed in order to find the eigenfunctions and eigenvalues of an electron in a 2D ring. One of the main consequences is the appearance of a geometry induced potential, which is given in terms of the Gaussian and the mean curvature. We have verified that the Gaussian curvature does not affect the electronic states of the system because we have assumed that its core is much smaller than the internal radius of the ring. Otherwise, the mean curvature adds a term to the effective angular moment. However, it was observed that only the energy of the state with $m=0$ was impacted with a substantial decrease when the magnetic field is null. The modifications due to the sample curvature are also observed in both the effective angular momentum and in the cyclotron frequency, through the quantum number $m$ and the magnetic field, respectively. The main physical consequence of this manifestation is the increase in the energies of the states. When Fermi energy is kept constant, we have checked that this last result implies in a decrease in the number of occupied states and, moreover, we have also checked that the subbands are more quickly depopulated. As a consequence, the width of the states decreases rapidly. We have also found that the curvature of the surface affects the conducting region of the ring, making it reach a maximum value at weak values of magnetic fields.

\section*{Acknowledgments}

This work was partially supported by the Brazilian agencies CAPES, CNPq,
FAPEMA and FAPPR. FMA acknowledges CNPq Grants 313274/2017-7 and
434134/2018-0, and FAPPR Grant 09/2016. EOS acknowledges CNPq Grants
427214/2016-5 and 303774/2016-9, and FAPEMA Grants 01852/14 and 01202/16.

\bibliography{References}

\end{document}